\documentclass[12pt]{iopart}
\usepackage{amssymb}
\usepackage{amsthm}
\usepackage{graphicx}
\usepackage{graphics,epsfig,color}

\newcommand{\bfx}{{\bf x}}
\newcommand{\bfxp}{{{\bf x}^\prime}}
\newcommand{\wbfx}{\widehat{\bf x}}
\newcommand{\wbfxp}{{\widehat{\bf x}^\prime}}
\newcommand{\N}{{\mathbf N}}

\newcommand{\R}{{\mathbf R}}
\newcommand{\Si}{{\mathbf S}}
\newcommand{\Hi}{{\mathbf H}}

\newcommand{\Z}{{\mathbf Z}}
\newcommand{\C}{{\mathbf C}}

\newcommand{\mcI}{{\mathcal I}}
\newcommand{\mch}{{\mathcal H}}
\newcommand{\mcg}{{\mathcal G}}

\newtheorem{lemma}{Lemma}[section]
\newtheorem{thm}[lemma]{Theorem}

\newtheorem{prop}[lemma]{Proposition}

\begin{document}

\title[
Fundamental solution of Laplace's equation in hyperbolic geometry]{Fundamental 
solution of the Laplacian in the hyperboloid model of hyperbolic geometry}

\author{H S Cohl${}^{1,2}$ and E G Kalnins${}^3$}

\address{$^1$Information Technology Laboratory, National Institute of Standards and Technology, Gaithersburg, MD, USA}
\address{$^2$Department of Mathematics, University of Auckland, Auckland, New Zealand}
\address{$^3$Department of Mathematics, University of Waikato, Hamilton, New Zealand}
\ead{hcohl@nist.gov}
\begin{abstract}
Due to the isotropy of $d$-dimensional hyperbolic space, one expects there to exist a spherically 
symmetric fundamental solution for its corresponding Laplace-Beltrami operator.  The $R$-radius 
hyperboloid model of hyperbolic geometry $\Hi_R^d$ with $R>0$, represents a Riemannian 
manifold with 
negative-constant sectional curvature.  We obtain a spherically symmetric fundamental solution of 
Laplace's equation on this manifold in terms of its geodesic radius.  We give several matching 
expressions for this fundamental solution including a definite integral over reciprocal powers of 
the hyperbolic sine, finite summation expression over hyperbolic functions, Gauss hypergeometric 
functions, and in terms of the associated Legendre function of the second kind with order and 
degree given by $d/2-1$ with real argument greater than unity.  We also demonstrate uniqueness
for a fundamental solution of Laplace's equation on this manifold in terms of a vanishing decay at infinity.
\end{abstract}

\maketitle

\section{Introduction}
\label{Introduction}

We compute closed-form expressions of a spherically symmetric Green's 
function (fundamental solution) of the Laplacian (Laplace-Beltrami operator)
on a Riemannian manifold of constant negative sectional curvature, namely 
the hyperboloid model of hyperbolic geometry.  Useful background material 
relevant for this paper can be found in Vilenkin (1968) \cite{Vilen}, 
Thurston (1997) \cite{Thurston}, Lee (1997) \cite{Lee} and 
Pogosyan \& Winternitz (2002) \cite{PogWin}.

This paper is organized as follows.  
In section \ref{Thehyperboloidmodelofhyperbolicgeometry} 
we describe the hyperboloid model of hyperbolic geometry and its corresponding metric, 
global geodesic distance function, Laplacian and geodesic polar coordinate systems
which parametrize points in this model.
In section \ref{AGreensfunctioninthehyperboloidmodel} for the hyperboloid model of hyperbolic
geometry, we show how to 
compute radial harmonics in a geodesic polar coordinate system and 
derive several alternative expressions for a radial fundamental solution of 
the Laplacian on the $d$-dimensional $R$-radius hyperboloid with $R>0$. 
In section \ref{Uniquenessoffundamentalsolutionintermsofdecayatinfinity} we 
prove that our derived fundamental solution is unique in terms of a vanishing
decay at infinity.

Throughout this paper we rely on the following definitions.  
For $a_1,a_2,\ldots\in\C$, if $i,j\in\Z$ and $j<i$ then
$\sum_{n=i}^{j}a_n=0$ and $\prod_{n=i}^ja_n=1$.
The set of natural numbers is given by $\N:=\{1,2,\ldots\}$, the set
$\N_0:=\{0,1,2,\ldots\}=\N\cup\{0\}$, and the set
$\Z:=\{0,\pm 1,\pm 2,\ldots\}.$  The set $\R$ represents the real numbers.

\section{The hyperboloid model of hyperbolic geometry}
\label{Thehyperboloidmodelofhyperbolicgeometry}

Hyperbolic space in $d$-dimensions is a fundamental example of a space 
exhibiting hyperbolic geometry.  It was developed independently by Lobachevsky 
and Bolyai around 1830 (see Trudeau (1987) \cite{Trudeau}).  It is a geometry analogous to 
Euclidean geometry, but such that Euclid's parallel postulate is 
no longer assumed to hold.

\begin{figure}[htbp!]
\begin{center}
\includegraphics[width=0.8\textwidth]{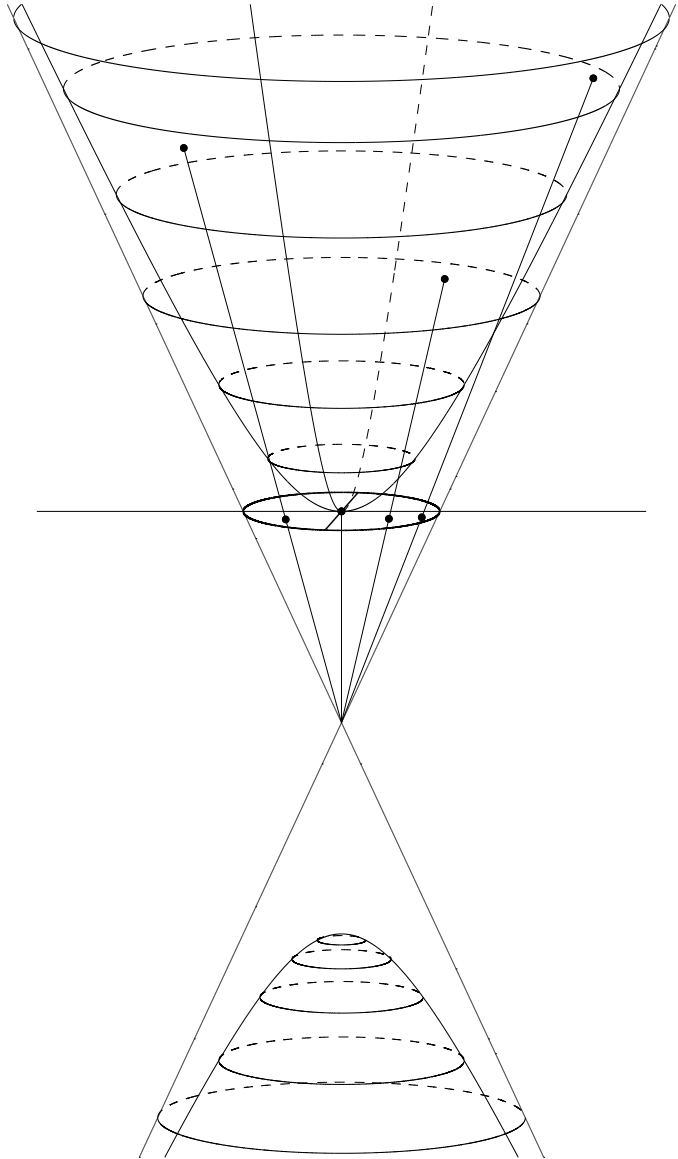}
\end{center}
\vspace{-1truecm}
\caption
{\rm This figure is a graphical depiction of stereographic projection from the hyperboloid model
to the Klein model of hyperbolic space.
}
\label{Fig:klein}
\end{figure}

\begin{figure}[htbp!]
\begin{center}
\includegraphics[width=0.8\textwidth]{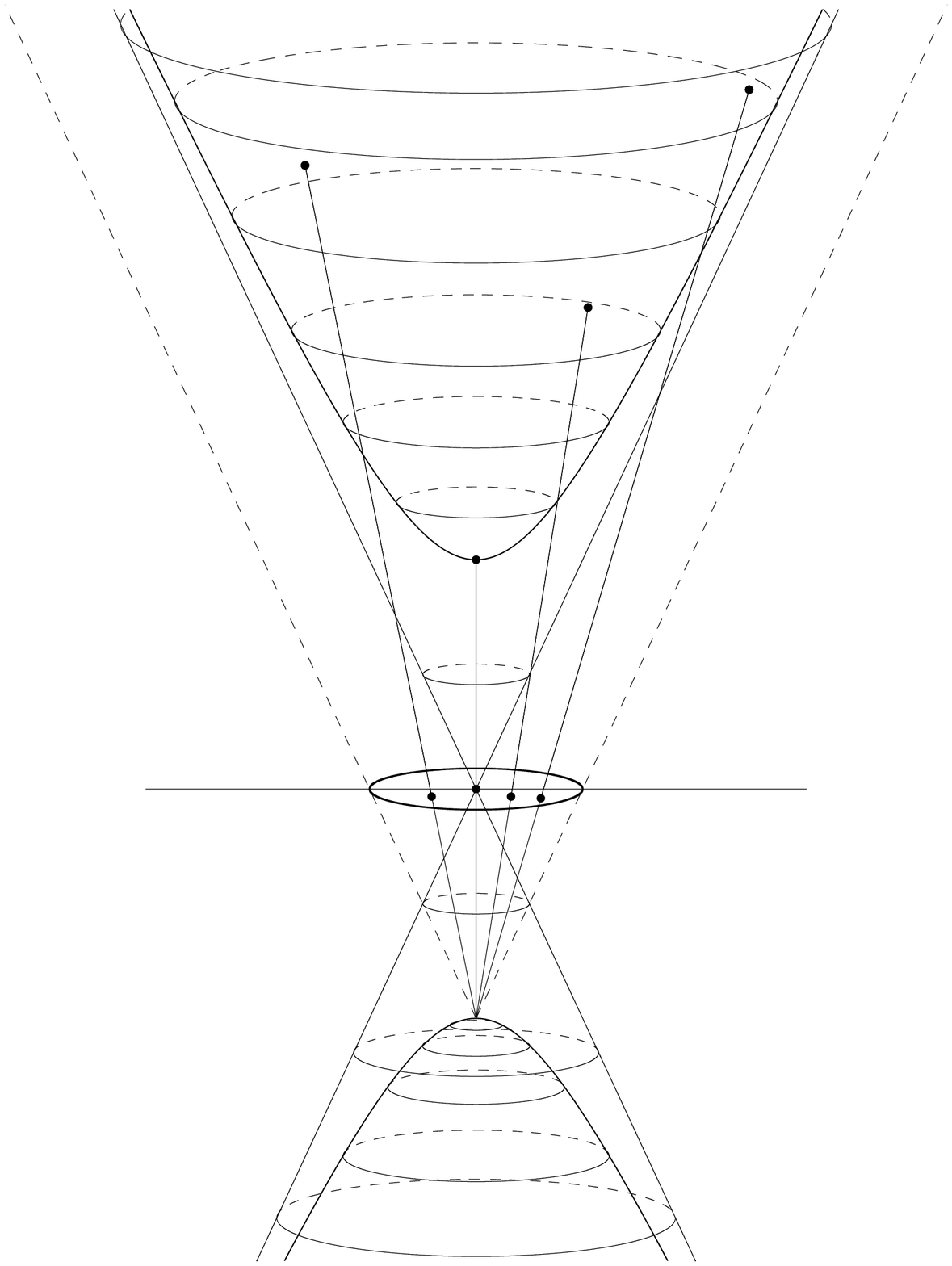}
\end{center}
\vspace{-1truecm}
\caption
{\rm This figure is a graphical depiction of stereographic projection from the hyperboloid model
to the Poincar\'{e} model of hyperbolic space.
}
\label{Fig:poin}
\end{figure}

There are several models of $d$-dimensional hyperbolic geometry including the 
Klein (see Figure \ref{Fig:klein}),
Poincar\'{e} (see Figure \ref{Fig:poin}),
hyperboloid, upper-half space and 
hemisphere models
(see Thurston (1997) \cite{Thurston}).  
The hyperboloid model for $d$-dimensional 
hyperbolic space is closely related to the Klein and Poincar\'{e}
models: each can be obtained projectively from the others.  
The upper-half space
and
hemisphere models can be obtained from one another by inversions with the 
Poincar\'{e} model
(see section 2.2 in Thurston (1997) \cite{Thurston}).
The model we will be focusing on 
in this paper is the hyperboloid
model.

The hyperboloid
model, also known as the Minkowski or Lorentz models, are models of 
$d$-dimensional hyperbolic geometry in which points are represented by 
the upper sheet (submanifold)
$S^+$ of a two-sheeted hyperboloid
embedded in 
the Minkowski space
$\R^{d,1}$.  Minkowski space is a
$(d+1)$-dimensional pseudo-Riemannian manifold
which is a real finite-dimensional vector space, with coordinates 
given by $\bfx=(x_0,x_1,\ldots,x_d)$. 
It is equipped with a nondegenerate, symmetric 
bilinear form, the Minkowski bilinear form
\[
[\bfx,{\mathbf y}]=x_0y_0-x_1y_1-\ldots-x_dy_d.
\]
The above bilinear form is symmetric, but not positive-definite, so it is 
not an inner product.  It is defined analogously with the Euclidean inner product 
for $\R^{d+1}$
\[
(\bfx,{\mathbf y})=x_0y_0+x_1y_1+\ldots+x_dy_d.
\]
The variety $[\bfx,\bfx]=x_0^2-x_1^2-\ldots-x_d^2=R^2$, for $\bfx\in\R^{d,1}$, using the language of 
Beltrami (1869) \cite{Beltrami} (see also p.~504 in Vilenkin (1968) \cite{Vilen}),
defines a pseudo-sphere of radius $R$.
Points on the pseudo-sphere with zero radius coincide with a cone.  
Points on the pseudo-sphere with radius greater than zero 
lie within this cone, and points on the pseudo-sphere
with purely imaginary radius lie outside the cone.

For $R\in(0,\infty)$, we refer to the variety $[\bfx,\bfx]=R^2$ as the 
$R$-radius hyperboloid $\Hi_R^d$.
This variety is a maximally symmetric, simply connected, $d$-dimensional Riemannian 
manifold with negative-constant sectional curvature (given by $-1/R^2$,
see for instance p.~148 in Lee (1997) \cite{Lee}),
whereas Euclidean space $\R^d$ equipped
with the Pythagorean norm, is a space with zero sectional curvature.
For a fixed $R\in(0,\infty),$ the $R$-radius hypersphere 
$\Si_R^d$, is an example of a space (submanifold) with positive 
constant sectional curvature (given by $1/R^2$).
We denote unit radius hyperboloid by $\Hi^d:=\Hi_1^d$ and 
the unit radius hypersphere by $\Si^d:=\Si_1^d$.

In our discussion of a fundamental solution for the Laplacian
in the hyperboloid model of hyperbolic geometry, we focus on the positive 
radius pseudo-sphere which can be parametrized through 
{\it subgroup-type coordinates,} i.e. those which correspond to a maximal
subgroup chain $O(d,1)\supset \ldots$ (see for instance Pogosyan \& Winternitz (2002) \cite{PogWin}).
There exist separable coordinate systems which parametrize points on the
positive radius pseudo-sphere (i.e. such as those which are analogous to
parabolic coordinates, etc.) which can not be constructed using maximal 
subgroup chains (we will no longer discuss these).

Geodesic polar coordinates are coordinates which correspond 
to the maximal subgroup chain given by $O(d,1)\supset O(d)\supset \ldots$.
What we will refer to as {\it standard geodesic polar coordinates} 
correspond to the subgroup chain given by 
$O(d,1)\supset O(d)\supset O(d-1)\supset \cdots \supset O(2).$
Standard geodesic polar coordinates
(see Olevski{\u\i} (1950) \cite{Olevskii};
Grosche, Pogosyan \& Sissakian (1997) \cite{groschepogsis}), similar 
to standard hyperspherical coordinates in Euclidean space, can be given by
\begin{equation}
\left.
\begin{array}{rcl}
x_0&=&R\cosh r\\[0.02cm]
x_1&=&R\sinh r\cos\theta_1\\[0.1cm]
x_2&=&R\sinh r\sin\theta_1\cos\theta_2\\[0.1cm]
&\vdots&\\[0.02cm]
x_{d-2}&=&R\sinh r\sin\theta_1\cdots\cos\theta_{d-2}\\[0.1cm]
x_{d-1}&=&R\sinh r\sin\theta_1\cdots\sin\theta_{d-2}\cos\phi\\[0.1cm]
x_{d}&=&R\sinh r\sin\theta_1\cdots\sin\theta_{d-2}\sin\phi,\\[0.1cm]
\end{array}
\quad\right\}
\label{standardhyp}
\end{equation}

\noindent where $r\in[0,\infty)$, $\phi\in[0,2\pi)$, 
and $\theta_i\in[0,\pi]$ for $i\in\{1,\ldots,d-2\}$.

The isometry group of the space $\Hi_R^d$ is the pseudo-orthogonal group $SO(d,1),$
the Lorentz group in $(d+1)$-dimensions.  Hyperbolic space $\Hi_R^d$, can be identified with 
the quotient space $SO(d,1)/SO(d)$.  The isometry group acts transitively on $\Hi_R^d$.  
That is, any point on the hyperboloid
can be carried, with the help of 
a Euclidean rotation of $SO(d-1)$, to the point 
$(\cosh\alpha,\sinh\alpha,0,\ldots,0),$ 
and a hyperbolic rotation
\begin{equation*}
\left.
\begin{array}{rcl}
x_0^\prime&=&-x_1\sinh\alpha+x_0\cosh\alpha\\[0.1cm]
x_1^\prime&=&-x_1\cosh\alpha-x_0\sinh\alpha\\[0.1cm]
\end{array}
\quad\right\}\nonumber
\end{equation*}
maps that point to the origin $(1,0,\ldots,0)$ of the space.
In order to study a fundamental solution of Laplace's equation on 
the hyperboloid, we need to describe how one computes distances in this space.  

One may naturally compare distances on the positive radius pseudo-sphere through 
analogy with the $R$-radius hypersphere.  Distances on the hypersphere
are simply given by arc lengths, angles between two 
arbitrary vectors, from the origin, in the ambient Euclidean space.
We consider the $d$-dimensional hypersphere embedded in $\R^{d+1}$.
Points on the hypersphere can be parametrized using 
hyperspherical coordinate systems.
Any parametrization of the hypersphere $\Si_R^d$, must have $(\bfx,\bfx)=x_0^2+\ldots+x_d^2=R^2$,
with $R>0$.  The distance between two points on the hypersphere $\bfx,\bfxp\in\Si_R^{d}$
is given by
\begin{equation}
\fl d(\bfx,\bfxp)=R\gamma
=R\cos^{-1}\left(\frac{(\bfx,\bfxp)}{(\bfx,\bfx)(\bfxp,\bfxp)} \right)
=R\cos^{-1}\left(\frac{1}{R^2}(\bfx,\bfxp)\right).
\label{cosgamma}
\end{equation}
This is evident from the fact that the geodesics on $\Si_R^d$ are great circles 
(i.e.~intersections of $\Si_R^d$ with planes through the origin) with constant speed 
parametrizations (see p.~82 in Lee (1997) \cite{Lee}).

Accordingly, we now look at the geodesic distance
function on the $d$-dimensional positive radius pseudo-sphere $\Hi_R^d.$ 
Distances between two points on the positive radius pseudo-sphere
are given by the hyperangle between two arbitrary vectors, from the origin, 
in the ambient Minkowski space.
\noindent Any parametrization of the hyperboloid $\Hi_R^d$, must have $[\bfx,\bfx]=R^2$.
The geodesic distance between two points $\bfx,\bfxp\in\Hi_R^d$ is given by
\begin{equation}
d(\bfx,\bfxp)=R\cosh^{-1}\left(\frac{[\bfx,\bfxp]}{[\bfx,\bfx][\bfxp,\bfxp]} \right)
=R\cosh^{-1}\left(\frac{1}{R^2}[\bfx,\bfxp]\right),
\label{dhyperboloid}
\end{equation}
where the inverse hyperbolic cosine with argument $x\in(1,\infty)$ is 
given by (see (4.37.19) in Olver {\it et al.} (2010) \cite{NIST})
\[
\cosh^{-1}x=\log\left(x+\sqrt{x^2-1}\right).
\]
Geodesics on $\Hi_R^d$ are great hyperbolas 
(i.e.~intersections of $\Hi_R^d$ with planes through the origin) with constant speed 
parametrizations (see p.~84 in Lee (1997) \cite{Lee}).
We also define a global function $\rho:\Hi^d\times\Hi^d\to[0,\infty)$
which represents the projection of global geodesic distance function
(\ref{dhyperboloid})
on $\Hi_R^d$ onto the corresponding unit radius hyperboloid $\Hi^d$, namely
\begin{equation}
\rho(\wbfx,\wbfxp):=d(\bfx,\bfxp)/R,
\label{rhodefn}
\end{equation}
where $\wbfx=\bfx/R$ and $\wbfxp=\bfxp/R$.

\subsection{The Laplacian on the hyperboloid model}

Parametrizations of a submanifold embedded in either a Euclidean or Minkowski space
is given in terms of coordinate systems whose coordinates are curvilinear.  These are 
coordinates based on some transformation that converts the standard Cartesian 
coordinates in the ambient space to a coordinate system with the same number of 
coordinates as the dimension of the submanifold in which the coordinate lines are curved.

On a $d$-dimensional Riemannian manifold $M$
(a manifold together with a Riemannian metric $g$), 
the Laplace-Beltrami operator (Laplacian) 
$\Delta:C^p(M)\to C^{p-2}(M),$ $p \ge 2$,
in curvilinear coordinates
${\mathbf{\xi}}=(\xi^1,\ldots,\xi^d)$ 
is given by
\begin{equation}
\Delta=\sum_{i,j=1}^d\frac{1}{\sqrt{|g|}}
\frac{\partial}{\partial \xi^i}
\left(\sqrt{|g|}g^{ij} 
\frac{\partial}{\partial \xi^j}
 \right),
\label{laplacebeltrami}
\end{equation}

\noindent where $|g|=|\det(g_{ij})|,$ the infinitesimal distance is given by
\begin{equation}
ds^2=\sum_{i,j=1}^{d}g_{ij}d\xi^id\xi^j,\ 
\label{metric}
\end{equation}

\noindent and
\[
\sum_{i=1}^{d}g_{ki}g^{ij}=\delta_k^j,
\]
where $\delta_i^j\in\{0,1\}$ is the Kronecker delta defined for all $i,j\in\Z$ such that
\begin{equation}
\delta_i^j:=
\left\{ \begin{array}{ll}
\displaystyle 1 &\qquad\mathrm{if}\ i=j, \nonumber \\[0.1cm]
\displaystyle 0 &\qquad\mathrm{if}\ i\ne j.
\end{array} \right.
\label{Kronecker}
\end{equation}
For a Riemannian submanifold, the relation between the metric 
tensor in the ambient space 
and $g_{ij}$ of (\ref{laplacebeltrami}) and (\ref{metric}) is
\[
g_{ij}({\mathbf{\xi}})=\sum_{k,l=0}^dG_{kl}\frac{\partial x^k}{\partial \xi^i}
\frac{\partial x^l}{\partial \xi^j}.
\]
On $\Hi_R^d$ the ambient space is Minkowski, and 
therefore $G_{ij}=\mathrm{diag}(1,-1,\ldots,-1)$.

The set of all geodesic polar coordinate systems on the hyperboloid correspond
to the many ways one can put coordinates on a hyperbolic hypersphere, i.e.,
the Riemannian submanifold $U\subset\Hi_R^d$ defined for a fixed $\bfxp\in\Hi_R^d$
such that $d(\bfx,\bfxp)=b=const,$ where $b\in(0,\infty)$.
These are coordinate systems which correspond to subgroup chains
starting with $O(d,1) \supset O(d) \supset \cdots$,
with standard geodesic polar coordinates given by (\ref{standardhyp}) 
being only one of them. (For a thorough description of these see section X.5 in 
Vilenkin (1968) \cite{Vilen}.)
They all share the property that they are described by 
$(d+1)$-variables: $r\in[0,\infty)$ plus $d$-angles each being given by the 
values $[0,2\pi)$, $[0,\pi]$, $[-\pi/2,\pi/2]$ or $[0,\pi/2]$ 
(see Izmest'ev {\it et al.}~(1999, 2001) \cite{IPSWa, IPSWb}). 

In any of the geodesic polar coordinate systems, the 
global geodesic distance between any two points on the hyperboloid 
is given by (cf.~(\ref{dhyperboloid})) 
\begin{equation}
d(\bfx,\bfxp)
=R\cosh^{-1}
\bigl( \cosh r\cosh r^\prime - \sinh r\sinh r^\prime\cos\gamma\bigr),
\label{globalgeodistgeodesic}
\end{equation}
where $\gamma$ is the unique separation angle
given in each 
hyperspherical coordinate system.
For instance, the separation angle in standard geodesic polar coordinates
(\ref{standardhyp})
is given by the formula
\begin{equation}
\displaystyle\cos\,\gamma=\cos(\phi-\phi^\prime)
\prod_{i=1}^{d-2}\sin\theta_i{\sin\theta_i}^\prime
+\sum_{i=1}^{d-2}\cos\theta_i{\cos\theta_i}^\prime
\prod_{j=1}^{i-1}\sin\theta_j{\sin\theta_j}^\prime.
\label{prodform}
\end{equation}
Corresponding separation angle formulae for any geodesic polar
coordinate system can be computed using (\ref{cosgamma}), (\ref{dhyperboloid}),
and the associated formulae for the appropriate inner-products.
Note that by making use of the isometry group $SO(d,1)$ to map 
$\bfxp$ to the origin, then $\rho=Rr$ for $\Hi_R^d$ and in particular $\rho=r$
for $\Hi^d$.  Hence, for the unit radius hyperboloid, there is no distinction
between the global geodesic distance and the $r$-parameter in a geodesic
polar coordinate system.  For the $R$-radius hyperboloid, the only distinction
between the global geodesic distance and the $r$-parameter is the multiplicative
constant $R$.

The infinitesimal distance in a geodesic polar coordinate system 
on this submanifold is given by
\begin{equation}
ds^2=R^2(dr^2+\sinh^2r\ d\gamma^2),
\label{stanhypmetric}
\end{equation}
where an appropriate expression for $\gamma$ in a curvilinear coordinate system is given.
If one combines 
(\ref{standardhyp}), 
(\ref{laplacebeltrami}), 
(\ref{prodform})
and (\ref{stanhypmetric}), then in a particular geodesic polar 
coordinate system, Laplace's equation
on $\Hi_R^d$ is given by
\begin{equation}
\Delta f=\frac{1}{R^2}\left[\frac{\partial^2f}{\partial r^2}
+(d-1)\coth r\frac{\partial f}{\partial r}
+\frac{1}{\sinh^2r} \Delta_{\Si^{d-1}}f\right]=0,
\label{genhyplap}
\end{equation}
where $\Delta_{\Si^{d-1}}$ is the corresponding Laplace-Beltrami operator
on the unit radius hypersphere $\Si^{d-1}$.

\section{A Green's function in the hyperboloid model}
\label{AGreensfunctioninthehyperboloidmodel}

\subsection{Harmonics in geodesic polar coordinates}
\label{SepVarStaHyp}

Geodesic polar coordinate systems 
partition the $R$-radius hyperboloid $\Hi_R^d$ into a family of $(d-1)$-dimensional 
hyperbolic hyperspheres, each with a radius $r\in(0,\infty),$ on which all possible 
hyperspherical coordinate systems for $\Si^{d-1}$ may be used
(see for instance Vilenkin (1968) \cite{Vilen}).  One then must also 
consider the limiting case for $r=0$ to fill out all of $\Hi_R^d$.
In geodesic polar coordinates one can compute the 
normalized hyperspherical harmonics in this space by solving the Laplace 
equation using separation of variables which results in a general procedure 
which is given explicitly in Izmest'ev {\it et al.}~(1999, 2001) \cite{IPSWa, IPSWb}. 
These angular harmonics are given as general expressions involving trigonometric 
functions, Gegenbauer polynomials and Jacobi polynomials.

The harmonics in geodesic polar coordinate systems
are given in terms of a radial solution multiplied by the 
angular harmonics.  The angular harmonics are eigenfunctions of the Laplace-Beltrami operator
on $\Si^{d-1}$ with unit radius which satisfy the following eigenvalue problem
\begin{equation*}
\Delta_{\Si^{d-1}}Y_l^K
(\wbfx)
=-l(l+d-2)Y_l^K
(\wbfx),
\end{equation*}
where 
$\wbfx\in\Si^{d-1}$,
$Y_l^K (\wbfx)$
are normalized hyperspherical harmonics,
$l\in\N_0$ is the 
angular momentum quantum number, and 
$K$ stands for the set of $(d-2)$-quantum numbers identifying 
degenerate harmonics for each $l$. 
The degeneracy 
\begin{equation*}
(2l+d-2)\frac{(d-3+l)!}{l!(d-2)!}
\end{equation*}
(see (9.2.11) in Vilenkin (1968) \cite{Vilen}),
tells you how many linearly independent solutions exist for a particular $l$ value and dimension $d$.
The hyperspherical harmonics are normalized such that
\[
\int_{\Si^{d-1}}Y_l^K(\wbfx)\overline{Y_{l^\prime}^{K^\prime}(\wbfx)}
d\omega=
\delta_{l}^{l^\prime}
\delta_{K}^{K^\prime},
\]
where $d\omega$ is the Riemannian (volume) measure (see for instance section 3.4 in
Grigor'yan (2009)\cite{Grigor}) on $\Si^{d-1}$ which is invariant under the 
isometry group $SO(d)$ (cf.~(\ref{eucsphmeasureinv})), and for $x+iy=z\in\C$, 
$\overline{z}=x-iy$, represents complex conjugation.
The generalized Kronecker delta $\delta_K^{K^\prime}$ 
(cf.~(\ref{Kronecker}))
is defined such that it equals 1 if 
all of the $(d-2)$-quantum numbers identifying degenerate harmonics for each $l$ coincide, 
and equals zero otherwise.

Since the angular solutions (hyperspherical harmonics) are well-known (see 
Chapter IX in Vilenkin (1968) \cite{Vilen}; 
Chapter 11 in Erd{\'e}lyi {\it et al.} (1981) \cite{ErdelyiHTFII}), 
we will now focus on the radial solutions on $\Hi_R^d$ in geodesic 
polar coordinates, which 
satisfy the following ordinary differential equation
(cf.~(\ref{genhyplap})) for all $R\in(0,\infty),$ namely
\[
\frac{d^2u}{dr^2}+(d-1)\coth r\frac{du}{dr}-\frac{l(l+d-2)}{\sinh^2r}u=0.
\]

\noindent Four solutions to this ordinary differential equation 
$u_{1\pm}^{d,l},u_{2\pm}^{d,l}:(1,\infty)\to\C$ are given by
\[
{\displaystyle u_{1\pm}^{d,l}(\cosh r)=\frac{1}{\sinh^{d/2-1}r}P_{d/2-1}^{\pm(d/2-1+l)}
(\cosh r)},
\]
\noindent and
\[
{\displaystyle u_{2\pm}^{d,l}(\cosh r)=\frac{1}{\sinh^{d/2-1}r}Q_{d/2-1}^{\pm(d/2-1+l)}
(\cosh r)},
\]
where $P_\nu^\mu,Q_\nu^\mu:(1,\infty)\to\C$ are associated Legendre functions of the first and
second kind respectively (see for instance Chapter 14 in Olver {\it et al.} (2010) \cite{NIST}).

Due to the fact that the space $\Hi_R^d$ is homogeneous with respect to 
its isometry group, the pseudo-orthogonal group $SO(d,1)$, and therefore 
an isotropic manifold, we expect that 
there exist a fundamental solution of Laplace's equation on this 
space with spherically symmetric dependence.
We specifically expect these solutions to be given in terms of associated 
Legendre functions of the second kind with argument given by $\cosh r$.  
This associated Legendre function naturally fits our requirements because 
it is singular at )$r=0$ and vanishes at infinity, whereas the associated 
Legendre functions of the first kind, with the same argument, are 
regular at $r=0$ and singular at infinity.

\subsection{Fundamental solution of the Laplacian}
\label{FunSolLapHd}

In computing a fundamental solution of the Laplacian on $\Hi_R^d$, we know that
\begin{equation*}
-\Delta \mch_R^d(\bfx,\bfxp) = \delta_g(\bfx,\bfxp),
\end{equation*}
where $g$ is the Riemannian metric on $\Hi_R^d$ and 
$\delta_g(\bfx,\bfxp)$ is the Dirac delta function on the manifold $\Hi_R^d$.  
The Dirac delta function
is defined for an open set $U\subset\Hi_R^d$ with $\bfx,\bfxp\in\Hi_R^d$ such that 
\[
\int_U\delta_g(\bfx,\bfxp) d\mathrm{vol}_g =
\left\{ \begin{array}{ll}
\displaystyle 1 &\qquad\mathrm{if}\ \bfxp\in U, \nonumber \\[0.1cm]
\displaystyle 0 &\qquad\mathrm{if}\ \bfxp\notin U,
\end{array} \right.
\]
where $d\mathrm{vol}_g$ is the Riemannian (volume) measure, invariant under the isometry group $SO(d,1)$ of 
the Riemannian manifold $\Hi_R^d$, given (in standard geodesic polar coordinates) by 
\begin{equation}
\fl d\mathrm{vol}_g=R^d
\sinh^{d-1}r\,dr\,d\omega:=
R^d\sinh^{d-1}r\,dr\,
\sin^{d-2}\theta_{d-1}\cdots\sin\theta_2 d\theta_{1}\cdots d\theta_{d-1}.
\label{eucsphmeasureinv}
\end{equation}
Notice that as $r\to 0^+$ that $d\mathrm{vol}_g$ goes to the Euclidean measure, invariant under the Euclidean 
motion group $E(d)$, in spherical coordinates.  Therefore
in spherical coordinates, we have the following
\[
\delta_g(\bfx,\bfxp)=\frac{\delta(r-r^\prime)}{R^d\sinh^{d-1}r^\prime}
\frac{\delta(\theta_1-\theta_1^\prime)\cdots\delta(\theta_{d-1}-\theta_{d-1}^\prime)}
{\sin\theta_2^\prime\cdots\sin^{d-2}\theta_{d-1}^\prime}.
\]

In general since we can add any harmonic function to a fundamental solution 
of the Laplacian and still have a fundamental solution, we will use 
this freedom to make our fundamental solution as simple as possible.  
It is reasonable to expect that there exists a particular spherically
symmetric fundamental solution
$\mch_R^d(\bfx,\bfxp)$ 
on the hyperboloid
with pure radial $\rho(\wbfx,\wbfxp):=d(\bfx,\bfxp)/R$
(cf.~(\ref{rhodefn})) and constant angular dependence 
(invariant under rotations centered about the origin), 
due to 
the influence of the point-like nature of the Dirac delta function.
For a spherically symmetric solution to the Laplace
equation,
the 
corresponding $\Delta_{\Si^{d-1}}$
term 
vanishes since only the $l=0$ term survives.
In other words, we expect there to exist a fundamental solution of Laplace's
equation such that
$\mch_R^d(\bfx,\bfxp)=f(\rho)$.

We have proven that on the $R$-radius hyperboloid $\Hi_R^d$, a Green's function for 
the Laplace operator (fundamental solution for the Laplacian) can be given as follows.
\begin{thm}
Let $d\in\{2,3,\ldots\}.$  Define $\mcI_d:(0,\infty)\to\R$ as
\[
\mcI_d(\rho):=\int_\rho^\infty\frac{dx}{\sinh^{d-1}x},
\]
$\bfx,\bfxp\in\Hi_R^d$, and 
$\mch_R^d:(\Hi_R^d\times\Hi_R^d)\setminus\{(\bfx,\bfx):\bfx\in\Hi_R^d\}\to\R$ 
defined such that
\[
\mch_R^d({\bf x},{\bf x}^\prime):=
{\displaystyle \frac{\Gamma\left(d/2\right)}{2\pi^{d/2}R^{d-2}}\mcI_d(\rho)},
\label{thmh1d}
\]
where $\rho:=\cosh^{-1}\left([\wbfx,\wbfxp]\right)$ is the geodesic distance 
between $\wbfx$ and $\wbfxp$ on the pseudo-sphere of unit radius $\Hi^d$,
with 
$\wbfx=\bfx/R,$ 
$\wbfxp=\bfxp/R$,
then $\mch_R^d$ is a fundamental solution for $-\Delta$
where $\Delta$ is the 
Laplace-Beltrami operator on $\Hi_R^d$.
Moreover,\\[-0.3cm]
\begin{eqnarray*}
\fl\mcI_d(\rho)=
\left\{ \begin{array}{ll}
\displaystyle (-1)^{d/2-1}\frac{(d-3)!!}{(d-2)!!}
\Biggl[\log\coth \frac{\rho}{2}
+\cosh \rho\sum_{k=1}^{d/2-1}\frac{(2k-2)!!(-1)^k}{(2k-1)!!\sinh^{2k}\rho}\Biggr]
&\hspace{-0.2cm}\mathrm{if}\  d\ \mathrm{even}, \\[0.6cm]
\left\{ \begin{array}{l}
\displaystyle
(-1)^{(d-1)/2}\Biggl[\frac{(d-3)!!}{(d-2)!!}\\[0.4cm]
\displaystyle \hspace{1.2cm}+\left(\frac{d-3}{2}\right)!
\sum_{k=1}^{(d-1)/2}
\frac{(-1)^k\coth^{2k-1}\rho}
{(2k-1)(k-1)!((d-2k-1)/2)!}\Biggr],
\\[0.45cm]
\mathrm{or} \\[0.0cm]
\displaystyle
(-1)^{(d-1)/2}\frac{(d-3)!!}{(d-2)!!}\left[1+\cosh\rho
\sum_{k=1}^{(d-1)/2}
\frac{(2k-3)!!(-1)^k}{(2k-2)!!\sinh^{2k-1}\rho}\right],
\end{array} \right\}  &\hspace{-0.2cm}\mathrm{if}\  d\ \mathrm{odd}.
\end{array} \right.\\[0.3cm]
\hspace{-1.47cm}
=\frac{1}{(d-1)\cosh^{d-1}\rho}\,
{}_2F_1\left(\frac{d-1}{2},\frac{d}{2};\frac{d+1}{2};\frac{1}{\cosh^2\rho}\right),\\[0.3cm]
\hspace{-1.47cm}
=\frac{1}{(d-1)\cosh \rho\,\sinh^{d-2}\rho}\,
{}_2F_1\left(\frac12,1;\frac{d+1}{2};\frac{1}{\cosh^2\rho}\right),\\[0.3cm]
\hspace{-1.47cm}
=\frac{e^{-i\pi(d/2-1)}}{2^{d/2-1}\Gamma\left(d/2\right)\sinh^{d/2-1}\rho}\,
Q_{d/2-1}^{d/2-1}(\cosh\rho),\\[0.3cm]
\end{eqnarray*}\\[-1.0cm]
where $!!$ is the double factorial, ${}_2F_1$ is the Gauss hypergeometric function,
 and  $Q_\nu^\mu$  is  the  associated  Legendre  function  of  the  second  kind.
\end{thm}

\medskip
In the rest of this section, we develop the material in order to prove this theorem.
\medskip

Due to the fact that the space $\Hi_R^d$ is homogeneous with respect to 
its isometry group $SO(d,1)$, and therefore an isotropic manifold, without loss
of generality, we are free to map the point $\bfxp\in\Hi_R^d$ to the origin.
In this case the global distance function $\rho:\Hi^d\times\Hi^d\to[0,\infty)$
coincides with the radial parameter in geodesic polar coordinates, and we may
interchange $r$ with $\rho$ accordingly
(cf.~(\ref{globalgeodistgeodesic}) with $r^\prime=0$) in our representation
of a fundamental solution of Laplace's equation on this manifold.
Since a spherically symmetric choice for a fundamental solution of Laplace's
equation is harmonic everywhere except at the origin, we may first set $g=f^\prime$ 
in (\ref{genhyplap}) and solve the first-order equation
\[
g^\prime+(d-1)\coth\rho\ g=0,
\]
which is integrable and clearly has the general solution 
\begin{equation}
g(\rho)=\frac{df}{d\rho}=c_0\sinh^{1-d}\rho,
\label{dfdr}
\end{equation}
where $c_0\in\R$ is a constant which depends on $d$.  
Now we integrate (\ref{dfdr}) to obtain a fundamental 
solution for the Laplacian in $\Hi_R^d$ 
\begin{equation}
\mch_R^d(\bfx,\bfxp)=c_0\mcI_d(\rho)+c_1,
\label{Hdid01}
\end{equation}
where 
\begin{equation}
\mcI_d(\rho):=\int_\rho^\infty\frac{dx}{\sinh^{d-1}x},
\label{In}
\end{equation}
and $c_0,c_1\in\R$ are constants which depend on $d$.
This definite integral result is mentioned in 
section II.5 of Helgason (1984) \cite{Helgason84} and as well in
Losev (1986) \cite{Losev}
.
Notice that we can add any harmonic function to 
(\ref{Hdid01})
and still have a fundamental solution of the Laplacian 
since a fundamental solution of the Laplacian must satisfy
\[
\int_{\Hi_R^d} (-\Delta\varphi)(\bfxp)\mch_R^d(\bfx,\bfxp)\ d\mathrm{vol}_g^\prime = \varphi(\bfx),
\]
for all $\varphi\in {\mathcal D}(\Hi_R^d),$ where ${\mathcal D}$ is the space of test functions,
and $d\mathrm{vol}_g^\prime$ is the Riemannian (volume) measure on $\Hi_R^d$, in the primed coordinates.
In particular, 
we notice that from our definition of
$\mcI_d$ (\ref{In}) that 
\begin{equation*}
\lim_{\rho\rightarrow\infty}\mcI_d(\rho)=0.
\end{equation*}
Therefore it is convenient to set $c_1=0$ leaving us with
\begin{equation}
\mch_R^d(\bfx,\bfxp)=c_0\mcI_d(\rho).
\label{Hdid}
\end{equation}

In Euclidean space $\R^d$, a Green's function for Laplace's 
equation (fundamental solution for the Laplacian) 
is well-known and is given in the following theorem
(see Folland (1976) \cite{Fol3}; p.~94, Gilbarg \& Trudinger (1983) \cite{GT}; 
p.~17, Bers {\it et al.}~(1964) \cite{BJS}, p.~211).
\begin{thm}
Let $d\in\N$. Define
\[
\mcg^d({\bf x},{\bf x}^\prime)=
\left\{ \begin{array}{ll}
\displaystyle\frac{\Gamma(d/2)}{2\pi^{d/2}(d-2)}\|{\bf x}-{\bf x}^\prime\|^{2-d}
& \qquad\mathrm{if}\ d=1\mathrm{\ or\ }d\ge 3,\\[10pt]
\displaystyle\frac{1}{2\pi}\log\|{\bf x}-{\bf x}^\prime\|^{-1}
& \qquad\mathrm{if}\  d=2, \nonumber
\end{array} \right. 
\]
then $\mcg_R^d$ is a fundamental solution 
for $-\Delta$
in Euclidean space $\R^d$,
where $\Delta$ is the Laplace operator in $\R^d$.
\label{thmg1n}
\end{thm}
\noindent Note most authors only present the above theorem for the case $d\ge 2$ but it is easily-verified to
also be valid for the case $d=1$ as well.  

The hyperboloid $\Hi_R^d$, being a manifold, must behave locally like 
Euclidean space $\R^d$.  Therefore for small $\rho$ we have 
$e^\rho\simeq 1+\rho$ and $e^{-\rho}\simeq 1-\rho$ and in that limiting regime 
\[
\mcI_d(\rho)\approx\int_\rho^1 \frac{dx}{x^{d-1}}\simeq
\left\{ \begin{array}{ll}
-\log \rho &\qquad\mathrm{if}\ d=2, \nonumber \\[0.2cm]
{\displaystyle \frac{1}{\rho^{d-2}}} &\qquad\mathrm{if}\ d\ge 3, \end{array} \right. 
\]
which has exactly the same singularity as a Euclidean fundamental solution
for Laplace's equation.  Therefore the proportionality constant $c_0$ is 
obtained by matching locally to a Euclidean fundamental solution of
Laplace's equation
\[
\mch_R^d=c_0 \mcI_d\simeq {\mathcal G}^d,
\]
near the singularity located at $\bfx=\bfxp$.

We have shown how to compute a fundamental solution of the 
Laplace-Beltrami operator on the hyperboloid
in terms of an improper integral
(\ref{In}).  We would now like to express this integral in terms of 
well-known special functions.
A fundamental solution $\mcI_d$ can be computed using elementary 
methods through its definition (\ref{In}).  In $d=2$ we have
\[
\mcI_2(\rho)=\int_\rho^\infty \frac{dx}{\sinh x}=
\frac{1}{2}\log \frac{\cosh \rho+1}{\cosh \rho-1}
=\log\coth\frac{\rho}{2},
\]
and in $d=3$ we have
\[
\mcI_3(\rho)=\int_\rho^\infty \frac{dx}{\sinh^2x}
=\frac{e^{-\rho}}{\sinh \rho}=\coth \rho-1.
\]
This exactly matches up to that given by (3.27) in Hostler (1955) \cite{Hostler}.
In $d\in\{4,5,6,7\}$ we have
\begin{eqnarray}
\mcI_4(\rho)&=&
-\frac12\log\coth\frac{\rho}{2}
+\frac{\cosh \rho}{2\sinh^2 \rho},\nonumber\\[0.15cm]
\mcI_5(\rho)&=&\frac13(\coth^3\rho-1)-(\coth \rho-1),\nonumber\\[0.15cm]
\mcI_6(\rho)&=&
\frac38\log\coth\frac{\rho}{2}
+\frac{\cosh \rho}{4\sinh^4\rho}
-\frac{3\cosh \rho}{8\sinh^2\rho},\quad\mathrm{and}\nonumber\\[0.15cm]
\mcI_7(\rho)&=&\frac15(\coth^5\rho-1)-\frac23(\coth^3\rho-1)+\coth \rho-1.\nonumber
\end{eqnarray}

Now we prove several equivalent finite summation expressions for $\mcI_d(\rho)$. We
wish to compute the antiderivative $\mathfrak{I}_m:(0,\infty)\to\R$, which is defined as
\[
\mathfrak{I}_m(x):=\int\frac{dx}{\sinh^mx},
\]
where $m\in\N$.  This antiderivative satisfies the following recurrence relation
\begin{equation}
\mathfrak{I}_m(x)=-\frac{\cosh x}{(m-1)\sinh^{m-1}x}-\frac{(m-2)}{(m-1)}\mathfrak{I}_{m-2}(x),
\label{antiderivreccurence}
\end{equation}
which follows from the identity
\[
\frac{1}{\sinh^{m}x}=
\frac{\cosh x}{\sinh^mx}\cosh x
-\frac{1}{\sinh^{m-2}x},
\]
and integration by parts.  The antiderivative $\mathfrak{I}_m(x)$ naturally breaks into
two separate classes, namely
\begin{eqnarray}
\fl\int\frac{dx}{\sinh^{2n+1}x}&=&(-1)^{n+1}\frac{(2n-1)!!}{(2n)!!}\nonumber\\[0.2cm]
&&\hspace{1.0cm}\times\left[\log\coth\frac{x}{2}+\cosh x\sum_{k=1}^n\frac{(2k-2)!!(-1)^k}{(2k-1)!!\sinh^{2k}x}\right]+C,
\label{antiderodd}
\end{eqnarray}
and
\begin{equation}
\fl\int\frac{dx}{\sinh^{2n}x}=
\left\{ \begin{array}{l}
\displaystyle
(-1)^{n+1}\frac{(2n-2)!!}{(2n-1)!!}\cosh x
\sum_{k=1}^n\frac{(2k-3)!!(-1)^k}{(2k-2)!!\sinh^{2k-1}x}+C, \qquad\mbox{or}\\[0.55cm]
\displaystyle
(-1)^{n+1}(n-1)!\sum_{k=1}^n\frac{(-1)^k\coth^{2k-1}x}{(2k-1)(k-1)!(n-k)!}+C,
\end{array} \right.
\label{antidereven}
\end{equation}
\noindent where $C$ is a constant.
The double factorial $(\cdot)!!:\{-1,0,1,\ldots\}\to\N$ is defined by
\[
n!!:=
\left\{ \begin{array}{ll}
\displaystyle n\cdot(n-2)\cdots 2 &\quad\mathrm{if}\ n\ \mathrm{even}\ge 2,
\nonumber \\[0.1cm]
\displaystyle n\cdot(n-2)\cdots 1 &\quad\mathrm{if}\ n\ \mathrm{odd}\ge 1,
\nonumber \\[0.1cm]
\displaystyle 1 &\quad\mathrm{if}\ n\in\{-1,0\}.
\nonumber
\end{array} \right.
\]
Note that $(2n)!!= 2^nn!$ for $n\in\N_0$.
The finite summation formulae for $\mathfrak{I}_m(x)$ all follow trivially by induction
using (\ref{antiderivreccurence}) and the binomial expansion (cf.~(1.2.2) in
Olver {\it et al.} (2010) \cite{NIST})
\[
(1-\coth^2x)^n=n!\sum_{k=0}^n\frac{(-1)^k\coth^{2k}x}{k!(n-k)!}.
\]
The formulae (\ref{antiderodd}) and (\ref{antidereven}) are essentially equivalent to
(2.416.2--3) in Gradshteyn \& Ryzhik (2007), except (2.416.3) is not defined for the
integrand $1/\sinh x$.   
By applying
the limits of integration from the definition of $\mcI_d(\rho)$ in (\ref{In}) to
(\ref{antiderodd}) and (\ref{antidereven}) we obtain the following finite summation
expressions for $\mcI_d(\rho)$
\begin{eqnarray}
&&\hspace{-3cm}\mcI_d(\rho)=
\left\{ \begin{array}{ll}
\displaystyle (-1)^{d/2-1}\frac{(d-3)!!}{(d-2)!!}\Biggl[\log\coth \frac{\rho}{2}
+\cosh \rho\sum_{k=1}^{d/2-1}\frac{(2k-2)!!(-1)^k}{(2k-1)!!\sinh^{2k}\rho}\Biggr]
&\hspace{-0.2cm}\mathrm{if}\  d\ \mathrm{even}, \\[0.6cm]
\left\{ \begin{array}{l}
\displaystyle
(-1)^{(d-1)/2}\Biggl[\frac{(d-3)!!}{(d-2)!!}\\[0.4cm]
\displaystyle \hspace{1.2cm}+\left(\frac{d-3}{2}\right)!
\sum_{k=1}^{(d-1)/2}
\frac{(-1)^k\coth^{2k-1}\rho}
{(2k-1)(k-1)!((d-2k-1)/2)!}\Biggr],
\\[0.45cm]
\mathrm{or} \\[0.0cm]
\displaystyle
(-1)^{(d-1)/2}\frac{(d-3)!!}{(d-2)!!}\left[1+\cosh\rho
\sum_{k=1}^{(d-1)/2}
\frac{(2k-3)!!(-1)^k}{(2k-2)!!\sinh^{2k-1}\rho}\right],
\end{array} \right\}  &\hspace{-0.2cm}\mathrm{if}\  d\ \mathrm{odd}.
\end{array} \right.\nonumber\\[0.2cm]
\displaystyle &&\hspace{-7cm}{}
\label{sumgradryzhikIn}
\end{eqnarray}


Moreover, the antiderivative (indefinite integral) can be given in terms of the Gauss
hypergeometric function 
\begin{equation}
\fl\int\frac{d\rho}{\sinh^{d-1}\rho}=\frac{-1}{(d-1)\cosh^{d-1}\rho}
\,{}_2F_1\left(\frac{d-1}{2},\frac{d}{2};\frac{d+1}{2};\frac{1}{\cosh^2\rho}\right)
+C,
\label{antiderivativecos2r}
\end{equation}
where $C\in\R$.
The Gauss hypergeometric function 
${}_2F_1:\C^2\times(\C\setminus-\N_0)\times\{z\in\C:|z|<1\}\to\C$
can be defined in terms of the infinite series
\[
{}_{2}F_1(a,b;c;z):=
\sum_{n=0}^\infty \frac{(a)_n(b)_n}{(c)_n n!}z^n
\]
(see (15.2.1) in Olver {\it et al.} (2010) \cite{NIST}),
and elsewhere in $z$ by analytic continuation. 
(see (2.1.5) in Andrews, Askey \& Roy 1999),
The Pochhammer symbol (rising factorial) $(\cdot)_{l}:\C\to\C$ is defined by
\[
(z)_n:=\prod_{i=1}^n(z+i-1),
\]
where $l\in\N_0$.  Note that
\[
(z)_{l}=\frac{\Gamma(z+{l})}{\Gamma(z)},
\]
for all $z\in\C\setminus-\N_0$.
The gamma function $\Gamma:\C\setminus-\N_0\to\C$
(see Chapter 5 in Olver {\it et al.} (2010) \cite{NIST}), which is
ubiquitous in special function theory, is an important combinatoric
function which generalizes the factorial function over the natural numbers.
It is naturally defined over the right-half complex plane through Euler's
integral (see (5.2.1) in Olver {\it et al.} (2010) \cite{NIST})
\[
\Gamma(z):=\int_0^\infty t^{z-1} e^{-t} dt,
\]
\noindent $\mbox{Re}\ \!z>0$.
Some properties of the gamma function, which we will find useful are included below.  
An important formula which the gamma function satisfies is the duplication
formula (i.e.,~(5.5.5) in Olver {\it et al.} (2010) \cite{NIST})
\begin{equation}
\Gamma(2z)=\frac{2^{2z-1}}{\sqrt{\pi}}\Gamma(z)\Gamma\left(z+\frac12\right),
\label{duplicationformulagamma}
\end{equation}
provided $2z\not\in-\N_0$,

The antiderivative (\ref{antiderivativecos2r}) is verified as follows.
By using
\[
\frac{d}{dz}{}_2F_1(a,b;c;z)=\frac{ab}{c}{}_2F_1(a+1,b+1;c+1;z)
\]
(see (15.5.1) in Olver {\it et al.} (2010) \cite{NIST}),
and the chain rule, we can show that 
\begin{eqnarray}
\hspace{-2.0cm}\frac{d}{d\rho}
\frac{-1}{(d-1)\cosh^{d-1}\rho}
\,{}_2F_1\left(\frac{d-1}{2},\frac{d}{2};\frac{d+1}{2};\frac{1}{\cosh^2\rho}\right)
&=&\nonumber\\[0.2cm]
&&\hspace{-6.5cm}\frac{\sinh\rho}{\cosh^d\rho}\,{}_2F_1\left(\frac{d-1}{2},\frac{d}{2};\frac{d+1}{2}
;\frac{1}{\cosh^2\rho}\right)\nonumber\\[0.2cm]
&&\hspace{-5.0cm}+\frac{d\sinh\rho}{(d+1)\cosh^{d+2}\rho}
\,{}_2F_1\left(\frac{d+1}{2},\frac{d+2}{2};\frac{d+3}{2}
;\frac{1}{\cosh^2\rho}\right).\nonumber
\end{eqnarray}
The second hypergeometric function can be simplified using Gauss' relations for contiguous
hypergeometric functions, namely
\[
\fl z\,{}_2F_1(a+1,b+1;c+1;z)=\frac{c}{a-b}\bigl[{}_2F_1(a,b+1;c;z)-{}_2F_1(a+1,b;c;z)\bigr]
\]
(see p.~58 in Erd{\'e}lyi {\it et al.}~(1981) \cite{Erdelyi}), and
\[
{}_2F_1(a,b+1;c;z)=\frac{b-a}{b}{}_2F_1(a,b;c;z)
+\frac{a}{b}\,{}_2F_1(a+1,b;c;z)
\]
(see (15.5.12) in Olver {\it et al.} (2010) \cite{NIST}).
By doing this, the term with the hypergeometric function cancels leaving only a term
which is proportional to a binomial through
\[
{}_2F_1(a,b;b;z)=(1-z)^{-a}
\]
(see (15.4.6) in Olver {\it et al.} (2010) \cite{NIST}),
which reduces to $1/\sinh^{d-1}\rho$.  By applying
the limits of integration from the definition of $\mcI_d(\rho)$ in (\ref{In}) to
(\ref{antiderivativecos2r}) we obtain the following Gauss hypergeometric representation
\begin{equation}
\mcI_d(\rho)=
\frac{1}{(d-1)\cosh^{d-1}\rho}
\,{}_2F_1\left(\frac{d-1}{2},\frac{d}{2};\frac{d+1}{2};\frac{1}{\cosh^2\rho}\right).
\label{Idthetagausscos}
\end{equation}
Using (\ref{Idthetagausscos}), we can write another expression for $\mcI_d(\rho)$.
Applying Eulers's transformation
\[
{}_2F_1(a,b;c;z)=(1-z)^{c-a-b}{}_2F_1\left(c-a,c-b;c;z\right)
\]
(see (2.2.7) in Andrews, Askey \& Roy (1999) \cite{AAR}),
to (\ref{Idthetagausscos}) produces
\[
\mcI_d(\rho)=
\frac{1}{(d-1)\cosh\rho\,\sinh^{d-2}\rho}
\,{}_2F_1\left(\frac12,1;\frac{d+1}{2};\frac{1}{\cosh^2\rho}\right).
\]

Our derivation for a fundamental solution of Laplace's equation on the $R$-radius hyperboloid
$\Hi_R^d$ in terms of the associated Legendre function of the second
kind is as follows.  By starting with (\ref{Idthetagausscos}) and the definition 
of the associated Legendre function of the second kind $Q_\nu^\mu:(1,\infty)\to\C,$  namely
\[
\fl Q_\nu^\mu(z):=\frac{\sqrt{\pi}e^{i\pi\mu}\Gamma(\nu+\mu+1)(z^2-1)^{\mu/2}}
{2^{\nu+1}\Gamma(\nu+\frac32)z^{\nu+\mu+1}}
\,{}_2F_1\left(
\frac{\nu+\mu+2}{2},
\frac{\nu+\mu+1}{2};
\nu+\frac32; \frac{1}{z^2}
\right),
\]
for $|z|>1$ and $\nu+\mu+1\notin-\N_0$ (see (8.1.3) in Abramowitz \& Stegun (1972) \cite{Abra}), 
we derive
\begin{equation}
\fl{}_2F_1\left(\frac{d-1}{2},\frac{d}{2};\frac{d+1}{2};\frac{1}{\cosh^2\rho}\right)=
\frac
{2^{d/2}\Gamma\left(\frac{d+1}{2}\right)\cosh^{d-1}\rho}
{\sqrt{\pi}e^{i\pi(d/2-1)}(d-2)!\sinh^{d/2-1}\rho}Q_{d/2-1}^{d/2-1}(\cosh\rho).
\label{repgausshypintermsoftheonewewant}
\end{equation}
We have therefore verified that the harmonics computed in 
section \ref{SepVarStaHyp}, namely $u_{2+}^{d,0},$ give an alternate form
of a fundamental solution for the Laplacian on the hyperboloid.
Using the duplication formula for gamma functions
(\ref{duplicationformulagamma}), (\ref{Idthetagausscos}), and
(\ref{repgausshypintermsoftheonewewant}), we derive
\[
\mcI_d(\rho)=
\frac{e^{-i\pi(d/2-1)}}{2^{d/2-1}\Gamma\left(d/2\right)\sinh^{d/2-1}\rho}
Q_{d/2-1}^{d/2-1}(\cosh\rho).
\]

Notice that our chosen fundamental solutions of the Laplacian on the
hyperboloid have the property that they tend towards zero at 
infinity (even for the $d=2$ case, unlike Euclidean fundamental solutions
of the Laplacian).  Therefore these Green's functions are positive 
(see Grigor'yan (1983) \cite{Grigor83};
Grigor'yan (1985) \cite{Grigor85}) and hence
$\Hi_R^d$ is not parabolic.
Note that as a result of our proof, we see that the relevant associated 
Legendre functions of the second kind for $d\in\{2,3,4,5,6,7\}$ are 
(cf.~(\ref{sumgradryzhikIn}))
\begin{eqnarray*}
Q_0(\cosh \rho)=\log\coth\frac{\rho}{2},\\
\frac{1}{\sinh^{1/2}\rho}Q_{1/2}^{1/2}(\cosh \rho)=i\sqrt{\frac{\pi}{2}}(\coth \rho-1),\\
\frac{1}{\sinh \rho}Q_1^1(\cosh \rho)=
\log\coth\frac{\rho}{2}-\frac{\cosh \rho}{\sinh^2 \rho},\\
\frac{1}{\sinh^{3/2}\rho}Q_{3/2}^{3/2}(\cosh \rho)=3i\sqrt{\frac{\pi}{2}}
\left(-\frac13\coth^3\rho+\coth \rho-\frac23\right), \\
\frac{1}{\sinh^2\rho}Q_2^2(\cosh \rho)=3\,\log\coth\frac{\rho}{2}-
2\frac{\cosh \rho}{\sinh^4\rho}-3\frac{\cosh \rho}{\sinh^2\rho},
\quad\ \mathrm{and}\\
\frac{1}{\sinh^{5/2}\rho}Q_{5/2}^{5/2}(\cosh \rho)=15i\sqrt{\frac{\pi}{2}}
\left(\frac{1}{15}\coth^5\rho-\frac23\coth^3\rho+\coth \rho-\frac{8}{15}\right).
\end{eqnarray*}

\medskip
The constant $c_0$ in a fundamental solution for the Laplace operator
on the hyperboloid
(\ref{Hdid}) is computed by locally matching up the singularity 
to a fundamental solution for the Laplace operator in Euclidean space, 
Theorem \ref{thmg1n}.  The coefficient $c_0$ depends on $d$. It is determined
as follows. For $d\ge 3$ we take the asymptotic expansion for 
$c_0\mcI_d(\rho)$
as $\rho$ approaches zero 
and match this to a fundamental solution of Laplace's equation for 
Euclidean space given in Theorem \ref{thmg1n}. This yields
\begin{equation}
\displaystyle c_0=\frac{\Gamma\left(d/2\right)}{2\pi^{d/2}}.
\label{c0gamma}
\end{equation}
For $d=2$ we take the asymptotic expansion for 
\[
c_0\mcI_2(\rho)=c_0\log\coth\frac{\rho}{2}\simeq
c_0\log\|\bfx-\bfxp\|^{-1}
\]
as $\rho$ approaches zero, and match this to 
$\displaystyle \mcg^2(\bfx,\bfxp)=(2\pi)^{-1}\log\|\bfx-\bfxp\|^{-1},$
therefore $\displaystyle c_0=(2\pi)^{-1}$.  This exactly
matches (\ref{c0gamma}) for $d=2$.  The 
derivation that $\mcI_d(\rho)$ is an fundamental solution of the Laplace
operator on the hyperboloid $\Hi^d_R$ and 
the functions for $\mcI_d(\rho)$ are computed above.  

The sectional curvature of a pseudo-sphere of radius $R$ is $-1/R^2$. 
Hence using results in Losev (1986) \cite{Losev}, all equivalent 
expressions in Theorem \ref{thmh1d} can be used for a fundamental solution of
the Laplace-Beltrami operator on the $R$-radius hyperboloid $\Hi_R^d$
(cf.~section \ref{Thehyperboloidmodelofhyperbolicgeometry}), namely (where $R$ is now a free parameter)
\[
\mch_R^d({\bf x},{\bf x}^\prime):=
{\displaystyle \frac{\Gamma\left(d/2\right)}{2\pi^{d/2}R^{d-2}}\mcI_d\left(\rho\right)}.
\]
The proof of Theorem \ref{thmh1d} is complete.

\medskip

Furthermore, due to a theorem proved in \cite{Losev}, all equivalent expressions for
$\mcI_d(\rho)$ in Theorem \ref{thmh1d} represent upper bounds for a fundamental 
solution of the Laplace-Beltrami operator on non-compact Riemannian manifolds 
with negative sectional curvature not exceeding $-1/R^2$ with $R>0$.\\[-0.2cm]

We would also like to mention that a similar computation for a fundamental solution
of Laplace's equation on the positive-constant sectional curvature compact manifold, 
the $R$-radius hypersphere, has recently been computed in 
Cohl (2011) \cite{Cohlhypsphfundsol}.

\section{Uniqueness of fundamental solution in terms of decay at infinity}
\label{Uniquenessoffundamentalsolutionintermsofdecayatinfinity}

\noindent It is clear that in general a fundamental solution of Laplace's equation in 
the hyperboloid model of hyperbolic geometry $\mch_R^d$ is 
not unique since one can add any
harmonic function $h:\Hi_R^d\to\R$ to $\mch_R^d$ and still obtain a solution to
\[
-\Delta \mch_R^d(\bfx,\bfxp) = \delta_g(\bfx,\bfxp),
\]
since $h$ is in the kernel of $-\Delta$.

\pagebreak[2]
\begin{prop}
There exists precisely one $C^\infty$-function 
$H:(\Hi_R^d\times\Hi_R^d)\setminus\{(\bfx,\bfx):\bfx\in\Hi_R^d\}\to\R$ such 
that for all $\bfxp\in\Hi_R^d$ the function
$H_\bfxp:\Hi_R^d\setminus\{\bfxp\}\to\R$ defined by $H_\bfxp(\bfx):=H(\bfx,\bfxp)$ 
is a distribution on $\Hi_R^d$ with 
\[
-\Delta H_\bfxp=\delta_g(\cdot,\bfxp)
\]
and
\begin{equation}
\lim_{d(\bfx,\bfxp)\to\infty}H_\bfxp(\bfx)=0,
\label{limitequationunique}
\end{equation}
\noindent where $d(\bfx,\bfxp)$ is the geodesic distance between two points $\bfx,\bfxp\in\Hi_R^d$.
\label{propuniquelaplace}
\end{prop}
\noindent Proof.  Existence: clear.  Uniqueness.  Suppose $H$ and $\tilde{H}$ are two 
such functions.  Let $\bfxp\in\Hi_R^d$.  Define the $C^\infty$-function 
$h:\Hi_R^d\setminus\{\bfxp\}\to\R$ by $h=H_\bfxp-\tilde{H}_\bfxp.$  Then $h$
is a distribution on $\Hi_R^d$ with $-\Delta h=0$.  
Since $\Hi_R^d$ is locally Euclidean one has by local elliptic regularity that
$h$ can be extended to a $C^\infty$-function $\hat{h}:\Hi_R^d\to\R$.  It follows
from (\ref{limitequationunique}) for $H$ and $\tilde{H}$ that 
\begin{equation}
\lim_{d(\bfx,\bfxp)\to\infty}\hat{h}(\bfx)=0.  
\label{limitequationunique2}
\end{equation}
The strong elliptic maximum/minimum principle on a Riemannian manifold for a bounded 
domain $\Omega$ states that if $u$ is harmonic, then the supremum/infimum of $u$ 
in $\Omega$ coincides with the supremum/infimum of $u$ on the boundary $\partial\Omega$.
By using a compact exhaustion sequence $\Omega_k$ in a non-compact connected Riemannian 
manifold and passing to a subsequence $\bfx_k\in\partial\Omega_k$ such that $\bfx_k\to\infty$,
the strong elliptic maximum/minimum principle can be extended to non-compact connected Riemannian 
manifolds with boundary conditions at infinity (see for instance section 8.3.2 in 
Grigor'yan (2009) \cite{Grigor}).
Taking $\Omega_k\subset\Hi_R^d$, the strong elliptic maximum/minimum principle for non-compact 
connected Riemannian manifolds implies using (\ref{limitequationunique2}) that $\hat{h}=0$.  Therefore
$h=0$ and $H(\bfx,\bfxp)=\tilde{H}(\bfx,\bfxp)$ for all $\bfx\in\Hi_R^d\setminus\{\bfxp\}$.\\

By Proposition \ref{propuniquelaplace}, for $d\ge 2$, the function $\mch_R^d$ is the
unique normalized fundamental solution of Laplace's equation which satisfies
the vanishing decay (\ref{limitequationunique}). \\[-0.2cm]

\section*{Acknowledgements}
Much thanks to Simon Marshall, A.~Rod Gover, Tom ter Elst, Shaun Cooper, George Pogosyan, 
Willard Miller, Jr., and Alexander Grigor'yan for valuable discussions.  
I would like to express my gratitude to Carlos Criado Camb\'{o}n in 
the Facultad de Ciencias at Universidad de M\'{a}laga for his assistance 
in describing the global geodesic distance function in the hyperboloid model.  
We would also like to acknowledge two anonymous referees whose comments
helped improve this paper.
I acknowledge funding for time to write this paper from the 
Dean of the Faculty of Science at the University of Auckland in the form of a three 
month stipend to enhance University of Auckland 2012 PBRF Performance.
Part of this work was conducted while H.~S.~Cohl was a National Research Council
Research Postdoctoral Associate in the Information Technology Laboratory at the 
National Institute of Standards and Technology, Gaithersburg, Maryland, U.S.A.

\section*{References}

\begin{thebibliography}{10}

\bibitem{Abra}
M.~Abramowitz and I.~A. Stegun.
\newblock {\em Handbook of mathematical functions with formulas, graphs, and
  mathematical tables}, volume~55 of {\em National Bureau of Standards Applied
  Mathematics Series}.
\newblock U.S. Government Printing Office, Washington, D.C., 1972.

\bibitem{AAR}
G.~E. Andrews, R.~Askey, and R.~Roy.
\newblock {\em Special functions}, volume~71 of {\em Encyclopedia of
  Mathematics and its Applications}.
\newblock Cambridge University Press, Cambridge, 1999.

\bibitem{Beltrami}
E.~Beltrami.
\newblock Essai d'interpr\'etation de la g\'eom\'etrie noneuclid\'eenne.
  {T}rad. par {J}. {H}o\"uel.
\newblock {\em Annales Scientifiques de l'\'Ecole Normale Sup\'erieure},
  6:251--288, 1869.

\bibitem{BJS}
L.~{Bers}, F.~{John}, and M.~{Schechter}.
\newblock {\em Partial differential equations}.
\newblock Interscience Publishers, New York, N.Y., 1964.

\bibitem{Cohlhypsphfundsol}
H.~S. Cohl.
\newblock Fundamental solution of laplace's equation in hyperspherical
  geometry.
\newblock {\em Symmetry, Integrability and Geometry: Methods and Applications},
  7(108), 2011.

\bibitem{Erdelyi}
A.~Erd{\'e}lyi, W.~Magnus, F.~Oberhettinger, and F.~G. Tricomi.
\newblock {\em Higher transcendental functions. {V}ol. {I}}.
\newblock Robert E. Krieger Publishing Co. Inc., Melbourne, Fla., 1981.

\bibitem{ErdelyiHTFII}
A.~Erd{\'e}lyi, W.~Magnus, F.~Oberhettinger, and F.~G. Tricomi.
\newblock {\em Higher transcendental functions. {V}ol. {II}}.
\newblock Robert E. Krieger Publishing Co. Inc., Melbourne, Fla., 1981.

\bibitem{Fol3}
G.~B. Folland.
\newblock {\em Introduction to partial differential equations}.
\newblock Number~17 in Mathematical Notes. Princeton University Press,
  Princeton, 1976.

\bibitem{GT}
D.~Gilbarg and N.~S. Trudinger.
\newblock {\em Elliptic partial differential equations of second order}.
\newblock Number 224 in Grundlehren der mathematischen Wissenschaften.
  Springer-Verlag, Berlin etc., second edition, 1983.

\bibitem{Grigor83}
A.~A. Grigor'yan.
\newblock Existence of the {G}reen function on a manifold.
\newblock {\em Russian Mathematical Surveys}, 38(1(229)):161--162, 1983.

\bibitem{Grigor85}
A.~A. Grigor'yan.
\newblock The existence of positive fundamental solutions of the {L}aplace
  equation on {R}iemannian manifolds.
\newblock {\em Matematicheskie Zametki}, 128(170)(3):354--363, 446, 1985.

\bibitem{Grigor}
A.~A. Grigor'yan.
\newblock {\em Heat kernel and analysis on manifolds}, volume~47 of {\em AMS/IP
  Studies in Advanced Mathematics}.
\newblock American Mathematical Society, Providence, RI, 2009.

\bibitem{groschepogsis}
C.~Grosche, G.~S. Pogosyan, and A.~N. Sissakian.
\newblock Path-integral approach for superintegrable potentials on the
  three-dimensional hyperboloid.
\newblock {\em Physics of Particles and Nuclei}, 28(5):486--519, 1997.

\bibitem{Helgason84}
S.~Helgason.
\newblock {\em Groups and geometric analysis: Integral geometry, invariant
  differential operators, and spherical functions}, volume 113 of {\em Pure and
  Applied Mathematics}.
\newblock Academic Press Inc., Orlando, FL, 1984.

\bibitem{Hostler}
L.~{Hostler}.
\newblock Vector spherical harmonics of the unit hyperboloid in {M}inkowski
  space.
\newblock {\em Journal of Mathematical Physics}, 18(12):2296--2307, 1977.

\bibitem{IPSWa}
A.~A. Izmest'ev, G.~S. Pogosyan, A.~N. Sissakian, and P.~Winternitz.
\newblock Contractions of {L}ie algebras and separation of variables. {T}he
  {$n$}-dimensional sphere.
\newblock {\em Journal of Mathematical Physics}, 40(3):1549--1573, 1999.

\bibitem{IPSWb}
A.~A. Izmest'ev, G.~S. Pogosyan, A.~N. Sissakian, and P.~Winternitz.
\newblock Contractions of {L}ie algebras and the separation of variables:
  interbase expansions.
\newblock {\em Journal of Physics A: Mathematical and General}, 34(3):521--554,
  2001.

\bibitem{Lee}
J.~M. Lee.
\newblock {\em Riemannian manifolds}, volume 176 of {\em Graduate Texts in
  Mathematics}.
\newblock Springer-Verlag, New York, 1997.

\bibitem{Losev}
A.~G. Losev.
\newblock Harmonic functions on manifolds of negative curvature.
\newblock {\em Matematicheskie Zametki}, 40(6):738--742, 829, 1986.

\bibitem{Olevskii}
M.~N. {Olevski{\u\i}}.
\newblock Triorthogonal systems in spaces of constant curvature in which the
  equation {$\Delta_2u+\lambda u=0$} allows a complete separation of variables.
\newblock {\em Matematicheski{\u\i} Sbornik}, 27(69):379--426, 1950.
\newblock (in Russian).

\bibitem{NIST}
F.~W.~J. Olver, D.~W. Lozier, R.~F. Boisvert, and C.~W. Clark, editors.
\newblock {\em N{IST} handbook of mathematical functions}.
\newblock Cambridge University Press, Cambridge, 2010.

\bibitem{PogWin}
G.~S. Pogosyan and P.~Winternitz.
\newblock Separation of variables and subgroup bases on {$n$}-dimensional
  hyperboloids.
\newblock {\em Journal of Mathematical Physics}, 43(6):3387--3410, 2002.

\bibitem{Thurston}
W.~P. {Thurston}.
\newblock {\em Three-dimensional geometry and topology. {V}ol. 1}, volume~35 of
  {\em Princeton Mathematical Series}.
\newblock Princeton University Press, Princeton, NJ, 1997.
\newblock Edited by Silvio Levy.

\bibitem{Trudeau}
R.~J. Trudeau.
\newblock {\em The non-Euclidean revolution}.
\newblock Birkh\"{a}user, Boston, 1987.

\bibitem{Vilen}
N.~Ja. Vilenkin.
\newblock {\em Special functions and the theory of group representations}.
\newblock Translated from the Russian by V. N. Singh. Translations of
  Mathematical Monographs, Vol. 22. American Mathematical Society, Providence,
  R. I., 1968.

\end{thebibliography}

\end{document}